\documentclass[manuscript, nonacm]{acmart}

\renewcommand\footnotetextcopyrightpermission[1]{} % Removes permission note
\setcopyright{none} % Disables ACM copyright
\acmConference{}{}{}
\acmBooktitle{}

\AtBeginDocument{%
  }

\usepackage{booktabs}
\usepackage{url}
\usepackage{amsmath,amssymb} % amssymb already loads amsfonts, so no clash
\usepackage{algorithm}
\usepackage{algorithmic}
\usepackage{subcaption}
\usepackage{algorithmic}
\usepackage{multirow}
\usepackage{graphicx}
\usepackage{textcomp}
\usepackage{xcolor}
\usepackage{svg}
\usepackage{enumitem}
\usepackage{pgfplots}
\usepackage{listings}
\usepackage{color}
\usepackage{tcolorbox}
\usepackage{enumitem}

\newcommand{\secref}[1]{Section~\ref{sec:#1}}
\newcommand{\tabref}[1]{Table~\ref{tab:#1}}

\newcommand{\figref}[1]{Figure~\ref{fig:#1}}

\newcommand{\SF}{LIA}
\newcommand{\DS}{DeepSeek-R1-Distill-Llama-8B}

\pgfplotsset{compat=1.17}

\begin{document}

\title{\SF{}: Supervised Fine-Tuning of Large Language Models for Automatic Issue Assignment}

\author{Arsham Khosravani}
\affiliation{%
  \institution{California State University Northridge}
  \city{Los Angeles}
  \state{California}
  \country{USA}}
\email{arsham.khosravani.085@my.csun.edu}

\author{Alireza Hoseinpour}
\affiliation{%
  \institution{Bowling Green State University}
  \city{Bowling Green}
  \state{Ohio}
  \country{USA}}
\email{ahosein@bgsu.edu}

\author{Arshia Akhavan}
\affiliation{%
  \institution{Bowling Green State University}
  \city{Bowling Green}
  \state{Ohio}
  \country{USA}}
\email{arshiaa@bgsu.edu}

\author{Mehdi Keshani}
\affiliation{%
  \institution{Bowling Green State University}
  \city{Bowling Green}
  \state{Ohio}
  \country{USA}}
\email{mkeshan@bgsu.edu}

\author{Abbas Heydarnoori}
\affiliation{%
  \institution{Bowling Green State University}
  \city{Bowling Green}
  \state{Ohio}
  \country{USA}}
\email{aheydar@bgsu.edu}

%%%%%%%%%%%%%%%%%%%%%%%%%%%%%%%%%%%%%%%%%
%%
%%          Abstract
%%
%%%%%%%%%%%%%%%%%%%%%%%%%%%%%%%%%%%%%%%%%
\begin{abstract}
Issue assignment is a critical process in software maintenance, where new issue reports are validated and assigned to suitable developers. However, manual issue assignment is often inconsistent and error-prone, especially in large open-source projects where thousands of new issues are reported monthly. Existing automated approaches have shown promise, but many rely heavily on large volumes of project-specific training data or relational information that is often sparse and noisy, which limits their effectiveness.
To address these challenges, we propose \emph{\SF{}} (\emph{LLM-based Issue Assignment}), which employs supervised fine-tuning to adapt an LLM, \DS{} in this work, for automatic issue assignment. By leveraging the LLM's pretrained semantic understanding of natural language and software-related text, \SF{} learns to generate ranked developer recommendations directly from issue titles and descriptions. The ranking is based on the model's learned understanding of historical issue-to-developer assignments, using patterns from past tasks to infer which developers are most likely to handle new issues. Through comprehensive evaluation, we show that \SF{} delivers substantial improvements over both its base pretrained model and state-of-the-art baselines. It achieves up to +187.8\% higher Hit@1 compared to the \DS{} pretrained base model, and outperforms four leading issue assignment methods by as much as +211.2\% in Hit@1 score. These results highlight the effectiveness of domain-adapted LLMs for software maintenance tasks and establish \SF{} as a practical, high-performing solution for issue assignment.
\end{abstract}


\maketitle

%%%%%%%%%%%%%%%%%%%%%%%%%%%%%%%%%%%%%%%%%
%%
%%          Section: Introduction
%%
%%%%%%%%%%%%%%%%%%%%%%%%%%%%%%%%%%%%%%%%%
\section{Introduction}\label{sec:intro}

In software development, issue tracking systems such as Jira, Bugzilla, and GitHub are used to manage and resolve issues reported by users or contributors. When a new issue is submitted, it is typically reviewed by a project maintainer or issue assigner, who verifies its validity, determines its severity, and assigns it to an appropriate developer~\cite{wu2022spatial, zhou2025issuecourier}. This process, known as \textit{issue assignment}, is essential for ensuring timely and accurate issue resolution. When issue assignment is ineffective, issues are often reassigned multiple times before reaching the right developer. For instance, in the Eclipse project, issue\# 16036~\cite{eclipsebug16036, eclipsebug16036activity} was reassigned more than a dozen times and took nearly 100 days to resolve, illustrating how manual assignment can be inefficient, increase maintenance costs, and delay software delivery~\cite{yadav2024developer, samir2023interpretable, wu2022spatial}.

Effective issue assignment is a challenging task. It requires detailed knowledge of the codebase, project structure, and the expertise and availability of contributors~\cite{hajari2024sofiawl, aini2025expertise}. These challenges are even more significant in open-source projects, where contributor roles are flexible and developer expertise is not always well documented. As a result, manual assignment can become inconsistent and error-prone~\cite{zhou2025issuecourier}. Furthermore, as software projects grow, the number of issue reports increases rapidly. Large-scale open-source projects like Eclipse, Mozilla, and Google Chromium receive thousands of new issue reports each month~\cite{xie2021devrec, zhou2025issuecourier}, making manual assignment increasingly unsustainable. This has led to the development of automated issue assignment solutions, which generally fall into two categories: \textit{text-based} and \textit{graph-based} methods~\cite{dong2024neighborhood}.

Text-based approaches frame issue assignment as a supervised learning task, where machine learning (ML) and deep learning (DL) techniques are used to predict the most suitable developer based on textual information such as issue titles and descriptions~\cite{dipongkor2023comparative, lee2022light, arnob2025bug, aung2022multi, mikolov2013efficient, mani2019deeptriage, wang2024empirical}. However, these methods come with notable limitations. They typically require large amounts of labeled data to perform well, which makes them less practical for smaller projects with limited issue histories~\cite{zhou2025issuecourier, arnob2025bug, akhavan2025linkanchorautonomousllmbasedagent, zhang2023ealink}. Such dependence on extensive labeled data reduces the generalizability of text-based methods and limits their effectiveness in real-world settings.

To overcome some of these challenges, state-of-the-art approaches have explored graph-based methods that represent relationships between developers, issue reports, and software artifacts~\cite{zhou2025issuecourier, cao2025complex, wu2022spatial, tao2025structural, dong2024neighborhood, dai2023graph}. These approaches often use graph neural networks (GNNs)~\cite{wu2020comprehensive} to learn embeddings that reflect structural, semantic, and temporal information. This helps model patterns such as developers fixing issues similar to ones they resolved in the past~\cite{wu2022spatial}. While graph-based methods can be effective, they are typically computationally expensive and rely on techniques like graph sampling or random walks~\cite{wu2022spatial}, which may miss important connections. They also require explicit links between issues and developers, which are often incomplete or noisy~\cite{dong2024neighborhood}.

Recent advancements in large language models (LLMs) have shown that they can achieve outstanding results on a variety of software maintenance tasks~\cite{aracena2025applying, izadi2022predicting, nikeghbal2024girt, xia2023automated, sarafraz2024domain, akhavan2025linkanchorautonomousllmbasedagent, Izadi:EMSE2021, yadollahi2025enhancing, abedini2025hybrid, Abedini:TOSEM2024}. This success is largely attributed to LLMs' pre-training on massive and diverse corpora, including software-related text and source code, which enables them to learn both general linguistic patterns and domain-specific programming knowledge~\cite{aracena2025applying}. As a result, LLMs offer a promising path to overcome two key limitations of prior issue assignment approaches: (i) the dependence of text-based methods on large labeled datasets, and (ii) the reliance of graph-based methods on complete and noise-free relational data, which is often impractical to construct in real-world settings. Furthermore, recent studies have shown that LLMs demonstrate stronger results on software maintenance tasks when they are domain-adapted through fine-tuning or instruction tuning~\cite{jin2023inferfix, aracena2025applying, zhang2023ealink, huang2025comprehensive, li2024exploratory, heo2025study}.

Motivated by these insights, We introduce an \emph{\underline{L}LM-based \underline{I}ssue \underline{A}ssignment} technique, called \emph{\SF{}}.
 We fine-tune an open-source LLM model, i.e., \DS{}~\cite{deepseekai2025deepseekr1incentivizingreasoningcapability}, using \emph{supervised fine-tuning}~\cite{radford2018improving, ouyang2022training,zhang2025instructiontuninglargelanguage} to help it learn the relationship between the issue report text and the developers who resolved them. We fine-tune the LLM for issue assignment using historical issue-assignee pairs. Each training instance consists of an issue title and description paired with its corresponding assignee. Through this process, the model learns associations between assignee identifiers and the semantic features of the issues they typically resolve. At inference time, when presented with unseen issues, the model leverages these learned patterns to generate a ranked list of candidate assignees.

We applied this fine-tuning to two widely used datasets in issue assignment research: EclipseJDT and Mozilla~\cite{bettenburg2008duplicate}. These datasets contain thousands of real-world issue reports along with the developers who fixed them. By learning from these examples, \SF{} can effectively recommend likely developers for new issue reports based solely on the issue text.

To evaluate the effectiveness of \SF{}, we conduct a comprehensive study comparing our fine-tuned model to both its pretrained base (\DS{}~\cite{deepseekai2025deepseekr1incentivizingreasoningcapability}) and to state-of-the-art issue assignment techniques. Specifically, we benchmark \SF{} and the base LLM against four representative approaches: NCGBT~\cite{dong2024neighborhood}, GCBT~\cite{dai2023graph}, GRCNN~\cite{wu2022spatial}, and CBR~\cite{anvik2006should}. We design our evaluation to answer two main research questions:

\begin{itemize}[leftmargin=18pt]
  \item \textbf{RQ1: How does \SF{} perform compared to the original pretrained model on the task of issue assignment?} \SF{} consistently outperformed the base model under zero-shot evaluation conditions across both datasets. The fine-tuned model achieved substantial gains in issue assignment accuracy, with improvements of up to +187.8\% at Hit@1 on EclipseJDT and +6.3\% at Hit@7 on Mozilla over its general-purpose pretrained base (i.e., \DS{}~\cite{deepseekai2025deepseekr1incentivizingreasoningcapability}).
  \item \textbf{RQ2: How does \SF{} compare against state-of-the-art approaches for issue assignment?} 
  \SF{} consistently surpassed all state-of-the-art baselines on both datasets. Compared to the strongest baseline, NCGBT, \SF{} achieved substantial improvements: up to +94.5\% at Hit@1 on EclipseJDT and an even higher gain of +211.2\% at Hit@1 on Mozilla.

\end{itemize}

In summary, our work makes the following key contributions:

\begin{itemize}[leftmargin=18pt]
    \item We present \SF{}, a fine-tuned LLM for issue assignment through supervised fine-tuning. By training on real-world issue reports with known assignees, \SF{} learns to map issue descriptions to the most relevant developers. This approach works well because it allows the model to internalize textual patterns from labeled examples, making it effective even in the absence of additional project metadata.
    
    \item We present a comprehensive empirical comparison of \SF{} against its base model and four state-of-the-art approaches.
\end{itemize}

The remainder of this paper is organized as follows. \secref{motivation} presents a motivating example from a real-world issue tracker to illustrate the challenges of manual issue assignment. \secref{method} describes the approach and implementation of \SF{}. \secref{eval} explains our evaluations and discusses the results. \secref{disc} discusses the implications of our findings. \secref{threats} addresses the potential threats to the validity of our findings. \secref{related} reviews related work. Finally, \secref{conc} concludes the paper and outlines future research directions.

%%%%%%%%%%%%%%%%%%%%%%%%%%%%%%%%%%%%%%%%%
%%
%%          Section: Motivation
%%
%%%%%%%%%%%%%%%%%%%%%%%%%%%%%%%%%%%%%%%%%

\section{Motivation}\label{sec:motivation}

\begin{figure*}[t]
  \centering
  \includegraphics[width=1.0\textwidth]{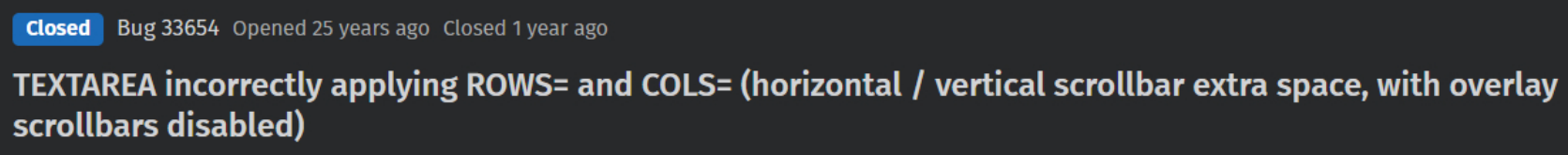}  
  \caption{Timeline of Mozilla issue 33654, which remained unresolved for nearly 25 years due to repeated reassignments, reopenings, and ownership uncertainty}
  \label{fig:bug life}
\end{figure*}

Manual issue assignment often introduces inefficiencies and delays in the software maintenance process. One particularly illustrative case is Eclipse issue 16036~\cite{eclipsebug16036, eclipsebug16036activity}, titled ``UI not responsive when stack is deep''. It shows a complex and extended reassignment pattern that exemplifies the practical challenges of manual developer assignment.

The issue was initially reported as new and quickly assigned to a developer, but within a short period, it was reassigned to another. Soon after, its status was changed back to new and then reassigned once more. Although it was marked as resolved with the resolution ``later,'' the issue was reopened several times over the following months, each time being passed between different developers. Its target milestone was also repeatedly modified, shifting across multiple release versions as the issue continued to resurface. This pattern of repeated reopening, reassignment, and milestone changes reflects the uncertainty and inefficiency that can occur when manual assignment fails to identify the most suitable developer early in the process.

A similar case can be found in Mozilla issue 33654~\cite{mozilla33654}, titled ``TEXTAREA incorrectly applying ROWS= and COLS= (horizontal / vertical scrollbar extra space, with overlay scrollbars disabled)'', whose lifespan is visualized in Figure~\ref{fig:bug life}. This issue remained open for nearly 25 years, offering a striking example of how inefficient assignment can delay even straightforward fixes. The issue was first reported in 1999 and immediately passed through multiple reassignments as project maintainers struggled to determine appropriate ownership. It was initially assigned and then reassigned several times in the following year, reflecting the uncertainty about which developer had the right expertise to address the problem.
Despite early attempts at resolution, including patches that led to the issue being marked as resolved, it was reopened multiple times when the issue resurfaced in testing. The report continued to change hands, with six different developers becoming involved over time. At several points, it was completely unassigned, reverting to Mozilla's default owner due to a lack of clarity or commitment. Even developers who began implementing partial fixes ultimately unassigned themselves, further delaying progress.
Over the next two decades, the issue saw little forward momentum. Its status shifted repeatedly between resolved and reopened, with no final fix accepted. This iterative cycle of reassignment and reopening exemplifies the challenges of manual issue assignment in large projects, where the absence of automated support and clear ownership can lead to persistent stagnation. In 2023, nearly after 25 years of its original filing, the issue was finally closed by marking it as a duplicate of a newer report that had implemented the necessary correction.

These cases highlight how manual assignment introduces significant inefficiencies into the issue assignment process. When issues are assigned to wrong developers, they often remain unresolved, resulting in unnecessary transfers and frequent reopenings. These delays, caused by unclear ownership and ineffective coordination, extend the time to resolution. Our work introduces \SF{}, a supervised fine-tuned~\cite{radford2018improving, ouyang2022training, zhang2025instructiontuninglargelanguage} LLM designed to automate issue assignment based on the textual content of issue reports. By training the model on historical issue data, \SF{} reduces the likelihood of initial misassignment and helps streamline the assignment process. This is especially valuable in large, collaborative projects where manual assignment may be inconsistent or overloaded.

%%%%%%%%%%%%%%%%%%%%%%%%%%%%%%%%%%%%%%%%%
%%
%%          Section: Approach
%%
%%%%%%%%%%%%%%%%%%%%%%%%%%%%%%%%%%%%%%%%%

\begin{figure*}[t]
  \centering
  \includegraphics[width=1.0\textwidth]{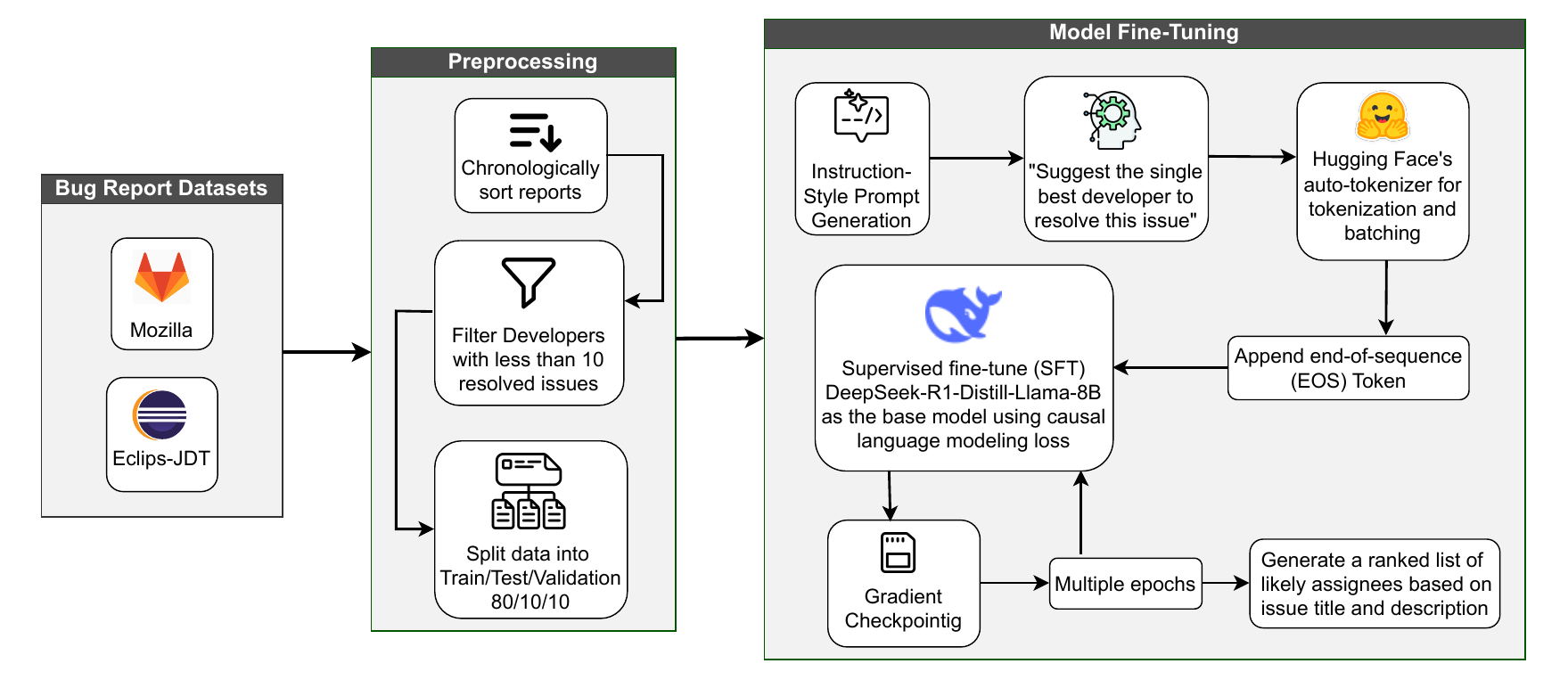}  %  
  \caption{Overview of \SF{}'s Architecture}
  \label{fig:overview}
\end{figure*}

\section{Approach}\label{sec:method}
In this section, we present a comprehensive overview of our methodology for adapting a pretrained LLM (\DS{}~\cite{deepseekai2025deepseekr1incentivizingreasoningcapability}) to the task of automated issue assignment. As illustrated in Figure~\ref{fig:overview}, our approach consists of two main stages: (1) data preparation and preprocessing and (2) supervised fine-tuning~\cite{radford2018improving, ouyang2022training,zhang2025instructiontuninglargelanguage} with instructional prompts. We begin by collecting and filtering issue tracker datasets from two large-scale open-source projects, EclipseJDT and Mozilla~\cite{bettenburg2008duplicate}. These reports are chronologically sorted, split into training and test sets, and formatted into instruction-style prompts. The model is fine-tuned using causal language modeling loss, learning to predict the most appropriate developer based solely on issue text. Through this process, the model learns from historical issue-to-developer assignments and uses those patterns to identify developers most likely to handle new issues based on similarity to past tasks.

\begin{figure}[t]
  \centering
  \begin{tcolorbox}[colback=white,colframe=black,boxrule=0.4pt,width=\linewidth]
  \textbf{Title}

  Google search engine has disappeared from search options, which makes Bing the default.

  \vspace{6pt}
  \textbf{Description}

  User Agent: \\
  Mozilla/5.0 (Android 8.1.0; Mobile; rv:66.0) \\
  Gecko/66.0 Firefox/66.0

  \vspace{4pt}
  \textit{Steps to reproduce:} \\
  Auto updated to 66.0a1. \\
  Searched for something (assumed using Google)

  \vspace{4pt}
  \textit{Actual results:} \\
  Bing results showed up. \\
  Bing had been made default, \\
  and Google completely disappeared from search options.

  \vspace{4pt}
  \textit{Expected results:} \\
  Search using Google.

  \vspace{6pt}
  \textbf{Assignee}

  \texttt{**Assignee Identifier**}
  \end{tcolorbox}
  \caption{Example issue report from the Mozilla dataset}
  \label{fig:example_issue}
\end{figure}

\subsection{Datasets}

We conducted our study using two large-scale datasets from established open-source software ecosystems, namely EclipseJDT and Mozilla~\cite{bettenburg2008duplicate}. Both projects maintain extensive public issue trackers, which made them ideal candidates for studying automated issue assignment in a practical and reproducible setting. These datasets have been widely used in prior work~\cite{dong2024neighborhood} and offer rich, mature histories of issue reports and assignee annotations, as illustrated in Figure~\ref{fig:example_issue}.

The EclipseJDT dataset included over 20,000 resolved issue reports. Each report contained metadata such as the issue title, detailed description, resolution status, priority level, and the identity of the assigned developer. The Mozilla dataset was significantly larger, comprising approximately 120,000 resolved issue reports, with a similar structure and level of detail.

Using this information, the model learns to map the textual content of an issue to the most likely developer assignment. By using these two distinct datasets, we ensured that our evaluation reflected different project dynamics, issue-reporting styles, and developer assignment distributions. EclipseJDT and Mozilla also varied considerably in size and structure, which provided a strong basis for comparing generalization and robustness across software ecosystems.

\subsection{Preprocessing}

To prevent temporal leakage, we adopted the same chronological partitioning strategy as NCGBT~\cite{dong2024neighborhood}, which was evaluated on the same datasets. Specifically, for each dataset, issue reports were sorted by their creation timestamps. Each dataset was partitioned into 80\% training, 10\% validation, and 10\% test portions. To ensure robustness and reduce potential bias from data ordering, we conducted multiple test runs in which the data was reshuffled and a different 10\% subset was used for evaluation in each run. The reported results represent the averaged performance across these runs, providing a more stable and reliable measure of the model's effectiveness.

Following NCGBT~\cite{dong2024neighborhood}, to ensure data quality and reduce label noise, we applied a filtering step that removed developers who had contributed to fewer than 10 resolved issues. All issue reports associated with these infrequent developers were excluded from the dataset. This threshold helped focus the learning process on consistently active contributors and reduced the likelihood of overfitting to sparsely represented classes.

After applying this filtering step, the final EclipseJDT dataset contained 16{,}106 resolved issue reports assigned to 4{,}017 developers. The Mozilla dataset, being significantly larger, included 110{,}467 resolved issue reports involving 37{,}371 developers. Each report retained key metadata such as the creation date, issue title, full description, and the identity of the developer responsible for its resolution.

These filtered datasets preserved a realistic and sufficiently large scale for both training and evaluation while also improving consistency by removing developers with minimal participation. This step reduced class sparsity and extreme imbalance, allowing the learning process to focus on stable patterns in developer assignment across both datasets.

For textual standardization, we represented each issue report by concatenating its title and description into a single prompt string. Developer identifiers were normalized using email addresses when available. We applied lowercasing strictly for matching during evaluation but retained the original casing in the model's output to preserve formatting fidelity.
A summary of the datasets after applying our preprocessing steps, including developer count, report volume and density, is shown in Table~\ref{tab:dataInfo}.

\begin{table}[t]
\centering
\caption{Summary of dataset characteristics and performance variation.}
\label{tab:dataInfo}
\begin{tabular}{lcc}
\toprule
\textbf{Factor} & \textbf{EclipseJDT} & \textbf{Mozilla} \\
\midrule
Developers & 4,017 & 37,371 \\
Issue Reports & 16,106 & 110,467 \\
Relationships & 53,985 & 569,289 \\
Density & 0.0008 & 0.0001 \\
\bottomrule
\end{tabular}
\end{table}

\subsection{Supervised Fine-Tuning with Instructional Prompts}

Supervised fine-tuning (SFT)~\cite{radford2018improving, ouyang2022training, zhang2025instructiontuninglargelanguage} is an effective method for adapting pretrained language models to specific tasks. Unlike general pretraining, where models learn broad language patterns from large amounts of unlabeled text, SFT focuses on teaching the model how to respond to specific types of inputs by training it on pairs of inputs and known outputs. In this setup, the model learns to generate correct responses when given structured instructions or queries, guided by examples that demonstrate the desired behavior.

In our project, we applied SFT to adapt the \DS{}~\cite{deepseekai2025deepseekr1incentivizingreasoningcapability} model for the task of issue assignment in software issue assignment. The goal was to train the model to predict the most appropriate developer to handle a given issue, using only its textual description. To do this, we fine-tuned the model on historical issue reports from EclipseJDT and Mozilla~\cite{bettenburg2008duplicate} datasets. They contain thousands of resolved issues annotated with their actual fixers.

Each issue report was transformed into an instruction-style prompt consisting of the issue title and description, followed by an ``Assignee'' line, as illustrated in Figure~\ref{fig:training_prompt}.
 This prompt simulated a realistic query from a human project maintainer. The model was trained to emit the correct developer identifier, immediately after the ``Assignee'' line. This identifier represented the actual developer who resolved the issue in historical records and served as the ground-truth label during training. We appended an end-of-sequence (EOS) token to each example, so the model would treat the task as a deterministic, single-span completion grounded in the context of the issue report.

The training objective used was causal language modeling loss, where the model learns to predict each next token in the sequence from left to right. In our case, this included both the natural language prompt and the expected developer label. This allowed the model to optimize its internal representations to maximize the likelihood of outputting the correct developer given an issue's text.

Our definition of the most appropriate or relevant developer is the one who historically resolved the given issue. While this does not guarantee that the same developer would be ideal for similar future issues, it provides a strong and practical training signal based on actual assignment decisions made by project maintainers. By learning from thousands of such mappings, the model implicitly captures patterns in how certain types of issues tend to be routed to specific developers.

\begin{figure}[t]
  \centering
  \begin{tcolorbox}[colback=white,colframe=black,boxrule=0.4pt,width=\linewidth]
  \textbf{ Prompt format used for instructing the model to predict the best developer for a given issue:} \\ 
  ``Below is a GitHub issue. Suggest the single best developer identifier to resolve it. Only return
  the identifier. 
  \texttt{Issue: + issue\_text + Assignee: **Assignee handler**''}
  \end{tcolorbox}
  \caption{Training Prompt Format}
  \label{fig:training_prompt}
\end{figure}

We preserved the issue content verbatim, with no manual edits, field normalization, or filtering. Prompts exceeding the model's 2048-token context window were truncated by removing tokens from the end of the issue report text, keeping the instruction and assignee. This choice ensured the task specification remained intact while maximizing the inclusion of meaningful context from the report.

\subsubsection{Tokenization}

Tokenization was handled using Hugging Face's AutoTokenizer, and batching was performed using data collator for language modeling, which applies standard causal language modeling loss.

Importantly, we did not mask out any tokens during training; the model was therefore optimized to predict every next token across the entire input, including both the prompt and the developer label. This full-sequence supervision worked well in practice and required no custom loss masking.

\subsubsection{Objective}
We trained the model using the standard causal language modeling loss, specifically left-to-right next-token prediction with cross-entropy, applied over the full concatenated sequence of {prompt, ground-truth assignee and EOS}. This setup encouraged the model to learn a deterministic, single-span completion behavior that maps issue reports to their corresponding developer identifiers. The format and supervision strategy remained consistent between training and evaluation under teacher-forcing.

\subsubsection{Training Configuration}
We fine-tuned~\DS{}, which has 8 billion parameters and uses a decoder-only architecture, under the following configuration:

\begin{itemize}[leftmargin=18pt]
    \item \textbf{Training Efficiency}: We used bfloat16 precision to reduce memory usage while maintaining numerical stability, and enabled TensorFloat-32 (TF32) operations on supported NVIDIA GPUs to accelerate matrix computations. Together, these optimizations improved training throughput and supported efficient scaling on large models without affecting accuracy.

    \item \textbf{Gradient checkpointing}: Enabled to reduce memory consumption during training by selectively storing intermediate activations. Instead of keeping all layer outputs in memory for backpropagation, this technique saves only a subset and recomputes during the backward pass. This trade-off between computation and memory allowed us to fit the model into available GPU memory without sacrificing model capacity or sequence length.
    \item \textbf{Disabling cache usage}: This setting was necessary to ensure compatibility with gradient checkpointing. By disabling the model's internal key-value caching mechanism during training, we avoided memory spikes and enabled stable backpropagation across long input sequences.
    \item \textbf{Optimizer}: AdamW~\cite{loshchilov2019decoupled} with weight decay set to 0.01; fused implementation used when supported.
    \item \textbf{Learning rate}: Set to a fixed value of $2 \times 10^{-5}$ with a warmup phase covering the first 3\% of training steps. During warmup, the learning rate increased linearly from zero to the target value, helping stabilize early training.
    \item \textbf{Batch size}: To simulate a larger batch size and stabilize gradient updates, we applied gradient accumulation over 4 steps. This resulted in an effective batch size of 4, meaning gradients were averaged across four forward-backward passes before each optimizer update.
    \item \textbf{Epochs}: The model was trained for 3 complete passes over each training dataset. Each epoch exposed the model to the full range of training examples, allowing it to iteratively refine its internal representations. This number of epochs was chosen to balance sufficient learning progress with overfitting risk, and we monitored training loss throughout to ensure convergence.
    \item \textbf{Context length}: The maximum input sequence length was set to 2048 tokens. This limit defined the amount of text the model could process in a single forward pass, including both the issue prompt and the expected developer identifier. Sequences longer than this threshold were truncated during preprocessing to fit within the context window, ensuring compatibility with the model's architecture while preserving as much relevant information as possible from the beginning of the input.
    \item \textbf{Reproducibility}: To ensure consistent and repeatable training behavior across runs, we fixed the random seed to 3407. This controlled sources of randomness such as weight initialization, data shuffling, and dropout patterns, enabling deterministic outcomes and reliable comparison of results across experiments.

\end{itemize}

We saved model checkpoints at regular intervals during training to preserve progress and allow for recovery if the process was interrupted. At the end of training, we saved a complete output folder containing all parts of the trained model and the tokenizer files, making the model ready for evaluation or future use. Training loss was tracked consistently to monitor learning progress and to identify any potential issues with model stability.

\subsubsection{Considerations for Long-Session Stability}

To support stable execution over extended training sessions, we applied several engineering safeguards:

\begin{itemize}[leftmargin=18pt]
  \item To ensure stable execution during multi-worker training and evaluation, we disabled parallel thread pools used by the tokenizer after process forking by setting the environment variable for tokenizer parallelism to false. This change helped prevent deadlocks that can occur when multiple threads are inherited across forked processes, particularly in environments like distributed training. Disabling tokenizer parallelism in this way contributed to smoother and more reliable pipeline behavior.
  \item We configured the PyTorch CUDA memory allocator to allow expandable memory segments, which helped reduce fragmentation in GPU memory during training. This was achieved by setting the internal allocation behavior to permit dynamically extendable segments, improving memory efficiency when working with large models and variable-length input sequences. This adjustment made memory usage more stable over time, especially during long training sessions where repeated allocations and deallocations could otherwise lead to wasted space or out-of-memory errors.
\end{itemize}

This combination of design choices, including the prompt structure, training objective, and tokenizer behavior, contributed to a training process that was stable, reproducible, and capable of handling large-scale issue assignment tasks. These design elements worked together to ensure that the model could effectively learn from real-world issue tracker data while maintaining performance and reliability across different experiment.

%%%%%%%%%%%%%%%%%%%%%%%%%%%%%%%%%%%%%%%%%
%%
%%          Section: Evaluation
%%
%%%%%%%%%%%%%%%%%%%%%%%%%%%%%%%%%%%%%%%%%
\section{Evaluation}\label{sec:eval}

In this section, we evaluate the effectiveness of our fine-tuned model, \SF{}, on the task of issue assignment. Specifically, we address the following research questions (RQs):

\begin{itemize}[leftmargin=18pt]
  \item \textbf{RQ1:} How does \SF{} perform compared to the original pretrained model on the task of issue assignment?

  \item \textbf{RQ2:} How does \SF{} compare against state-of-the-art specialized approaches for issue assignment?

\end{itemize}

\subsection{Evaluation Metric}

We formulate issue assignment as a developer recommendation task framed as a top-$K$ ranking problem. Let $\mathcal{B}_{\text{test}}=\{b_1,\ldots,b_N\}$ be the test issue reports and $\mathcal{D}$ the set of developers. Each $b_i$ has combined text $T_i$ (title and description) and a ground-truth developer $d_i^* \in \mathcal{D}$. Given $T_i$, the model produces a ranked list of candidates $R_i=(d_{i,1},\ldots,d_{i,K_{\max}})$; we denote its top-$K$ prefix by $R_i^{(K)}=\{d_{i,1},\ldots,d_{i,K}\}$.

Following prior work~\cite{dong2024neighborhood}, we evaluate with Hit@K, the proportion of test cases in which the true developer appears in the top-$K$:
\begin{equation}
\text{Hit@K} \;=\; \frac{1}{N} \sum_{i=1}^{N} \mathbf{1}\!\left\{\, d_i^* \in R_i^{(K)} \,\right\}.
\label{eq:HitAtK}
\end{equation}

Hit@K is a practical measure of ranking accuracy for issue assignment. We report results for $K \in \{1,\ldots,10\}$ to capture performance under different recommendation cutoffs.

\subsection{Experimental Setup}

To ensure reproducibility and a fair comparison between models, we used a structured and deterministic prompting procedure during evaluation. The key steps are:

\begin{enumerate}[leftmargin=18pt]
  \item \textbf{Constrained candidate space:} For each project, we first constructed a fixed candidate set \(\mathcal{C}\), consisting of all unique assignee handlers. During inference, we accepted only those model predictions that exactly matched a handler from this set. This restriction ensured that all predicted developers were real individuals from the dataset and prevented the model from generating hallucinated or fabricated identifiers that were not grounded in the task data.

  \item \textbf{Evaluation prompt:} For every test issue, we generated a structured prompt that asked the model to return a ranked list of developer email addresses. The prompt followed a consistent template that instructed the model to list exactly \(10\) known developers in descending order of suitability, using only handlers observed in the dataset. The evaluation prompt format is illustrated in Figure~\ref{fig:evaluation_prompt}.
  This clear and strict instruction helped constrain the output space and guided the model toward producing valid and well-structured responses.

  \item \textbf{Greedy decoding and postprocessing:} During generation, we decoded outputs using greedy decoding, meaning we selected the highest-probability token at each step. After decoding, we postprocessed the output by extracting only those tokens that matched valid email addresses in the candidate set. Any malformed or unrecognized entries were filtered out.

\end{enumerate}

To reduce the impact of variability in model generation and ensure stable evaluation outcomes, each experiment was repeated three times, and the average performance was reported.

\begin{figure}[t]
  \centering
  \begin{tcolorbox}[colback=white,colframe=black,boxrule=0.4pt,width=\linewidth]
  \textbf{ Prompt format used for instructing the model to list exactly 10 known developers:} \\ 
  ``Below is a GitHub issue. List the TOP 10 developers to handle the issue, ranked from best to worst.
  Use only developer identifiers known in this project. Return EXACTLY 10
  comma-separated items, unique, with no extra text.
  \texttt{Issue: + issue\_text + Top 10 assignees:}''
  \end{tcolorbox}
  \caption{Evaluation Prompt Format}
  \label{fig:evaluation_prompt}
\end{figure}

\subsection{Evaluation Baselines}

\subsubsection{\textbf{Baseline for RQ1}}
We use the original~\DS{} model~\cite{deepseekai2025deepseekr1incentivizingreasoningcapability} as our baseline. 
This model, trained on general web text without any task-specific fine-tuning, serves as a strong zero-shot reference point to assess the effectiveness of domain adaptation for issue assignment.

\subsubsection{\textbf{Baselines for RQ2}}
We selected four state-of-the-art methods for comparison.

\begin{itemize}[leftmargin=18pt]
    \item \textbf{NCGBT}~\cite{dong2024neighborhood}: A graph-based model designed for issue assignment that uses graph neural networks to model relationships from issue reports and developer collaboration networks for improved recommendations.
    \item \textbf{GCBT}~\cite{dai2023graph}: A graph-based collaborative filtering model that constructs a heterogeneous graph of issue and developers. It initializes issue nodes using a GRU-based text encoder, then applies a graph neural network to learn representations of both issues and developers. Predictions are made using an information retrieval strategy.
    \item \textbf{GRCNN}~\cite{wu2022spatial}: This model uses an LSTM-based module for issue reports and separately models developer relationships as a graph. It applies random walks to generate developer sequences from the graph and combines them with the textual information for final prediction.
    \item \textbf{CBR}~\cite{anvik2006should}: This text-based method uses a bag of words based on word frequency to convert the issue description into a vector, and then classifies it with SVM.
\end{itemize}

\subsection{Evaluation Results}

\subsubsection{\textbf{RQ1: How does \SF{} perform compared to the original pretrained model (Base) on the task of issue assignment across two real-world issue report datasets?}}

\begin{table}[t]
\centering
\caption{Hit@K results comparing \SF{} with Base on EclipseJDT and Mozilla (RQ1).}
\label{tab:rq1}
\begin{tabular}{@{}lcccccc@{}}
\toprule
\multirow{2}{*}{$K$} & \multicolumn{3}{c}{Eclipse-JDT} & \multicolumn{3}{c}{Mozilla} \\
\cmidrule(lr){2-4} \cmidrule(lr){5-7}
& Base & \SF{} & Improve & Base & \SF{} & Improve  \\
\midrule
1  & 0.156 & \textbf{0.449} & +187.8\% & 0.730 & \textbf{0.733} & +0.4\% \\
2  & 0.264 & \textbf{0.559} & +111.7\% & 0.745 & \textbf{0.777} & +4.3\% \\
3  & 0.344 & \textbf{0.620} & +80.2\% & 0.758 & \textbf{0.799} & +5.4\% \\
4  & 0.419 & \textbf{0.660} & +57.5\% & 0.766 & \textbf{0.810} & +5.7\% \\
5  & 0.475 & \textbf{0.684} & +44.0\% & 0.771 & \textbf{0.816} & +5.8\% \\
6  & 0.523 & \textbf{0.702} & +34.2\% & 0.776 & \textbf{0.824} & +6.2\% \\
7  & 0.569 & \textbf{0.713} & +25.3\% & 0.781 & \textbf{0.830} & +6.3\% \\
8  & 0.601 & \textbf{0.725} & +20.6\% & 0.784 & \textbf{0.833} & +6.2\% \\
9  & 0.633 & \textbf{0.749} & +18.3\% & 0.789 & \textbf{0.836} & +5.9\% \\
10 & 0.661 & \textbf{0.774} & +17.1\% & 0.794 & \textbf{0.839} & +5.7\% \\
\bottomrule
\end{tabular}
\end{table}

\tabref{rq1} reports the performance of \SF{} compared to the base pretrained model on EclipseJDT and Mozilla. Across both datasets, \SF{} consistently outperformed the baseline in Hit@k at all values of $K$, demonstrating the benefits of supervised fine-tuning for adapting an open-source LLM to issue assignment.

On EclipseJDT, the gains are especially pronounced. The base model struggled on this dataset, but fine-tuning led to large improvements, ranging from +17.1\% at Hit@10 up to +187.8\% at Hit@1. Moreover, the improvements followed a clear trend: the smaller the value of $K$, the larger the relative gain. For example, the model achieved +111.7\% at Hit@2, +80.2\% at Hit@3, and +57.5\% at Hit@4. The gains remained substantial beyond that as well, with +44.0\% at Hit@5, +34.2\% at Hit@6, and +25.3\% at Hit@7. This suggests that fine-tuning was particularly effective at boosting the model's top-ranked predictions.

On Mozilla, the base model already showed strong performance. Even so, \SF{} achieved consistent improvements, with the largest relative gain of +6.3\% at Hit@7. Gains remained steady across other $K$ values, including +5.8\% at Hit@5, +6.2\% at Hit@6, and +6.2\% at Hit@8, though they became smaller as $K$ decreased, reaching just +0.4\% at Hit@1. This suggests that while fine-tuning still added value, the improvement margin was narrower at lower rank positions when the base model was already performing well.

These differences in improvement reflect key characteristics of the datasets. As shown in Table~\ref{tab:dataInfo}, EclipseJDT contains a smaller, more balanced developer set and a higher density of issue-to-developer relationships, which allows the model to observe more consistent patterns during fine-tuning. Mozilla, in contrast, has a much larger and more imbalanced developer pool, with sparser supervision per developer. While Mozilla provides a larger volume of data overall, its higher variance and lower density make it more challenging for the model to learn strong associations between developers and the semantic characteristics of the issues they typically resolve.

The performance of the base model before fine-tuning also plays a role. On EclipseJDT, the pretrained model performed relatively poorly, leaving greater room for improvement. On Mozilla, where the base model already performed well, the potential gains were naturally more limited. Ultimately, the degree of improvement depends on both dataset structure and how well the base model aligns with the data. In all cases, fine-tuning enhanced assignment accuracy.

\begin{tcolorbox}
\textbf{Answer to RQ1:} \SF{} consistently outperformed the base model on both datasets, achieving improvements of up to +187.8\% (Hit@1) on EclipseJDT and +6.3\% (Hit@7) on Mozilla.
\end{tcolorbox}

\subsubsection{\textbf{RQ2: How does \SF{} compare against state-of-the-art approaches for issue assignment?}}

\begin{table*}[t]
\centering
\caption{Hit@K results comparing \SF{} with state-of-the-art graph-based methods on EclipseJDT and Mozilla (RQ2). ``Improve'' shows \SF{}'s percentage gain over NCGBT, which is the top-performing baseline.}
\label{tab:rq2}
\begin{tabular}{@{}lcccccc|cccccc@{}}
\toprule
\multirow{2}{*}{$K$} 
& \multicolumn{6}{c|}{EclipseJDT} 
& \multicolumn{6}{c}{Mozilla} \\
\cmidrule(lr){2-7} \cmidrule(lr){8-13}
& \SF{} & NCGBT & GCBT & GRCNN & CBR & Improve & \SF{} & NCGBT & GCBT & GRCNN & CBR & Improve \\
\midrule
1  & \textbf{0.449} & 0.2307 & 0.2018 & 0.1903 & 0.1356 & +94.5\% & \textbf{0.733} & 0.2356 & 0.2149 & 0.1925 & 0.1207 & +211.2\% \\
2  & \textbf{0.559} & 0.3406 & 0.2933 & 0.2102 & 0.1668 & +64.2\% & \textbf{0.777} & 0.2964 & 0.2658 & 0.2488 & 0.2031 & +162.1\% \\
3  & \textbf{0.620} & 0.4122 & 0.3604 & 0.3223 & 0.2377 & +50.5\% & \textbf{0.799} & 0.3618 & 0.3288 & 0.2874 & 0.2200 & +120.9\% \\
4  & \textbf{0.660} & 0.4721 & 0.4275 & 0.3456 & 0.2551 & +39.8\% & \textbf{0.810} & 0.4085 & 0.3670 & 0.3120 & 0.2699 & +98.3\% \\
5  & \textbf{0.684} & 0.5121 & 0.4752 & 0.3654 & 0.2670 & +33.6\% & \textbf{0.816} & 0.4085 & 0.3670 & 0.3120 & 0.2832 & +99.8\% \\
6  & \textbf{0.702} & 0.5592 & 0.5096 & 0.3907 & 0.3296 & +25.6\% & \textbf{0.824} & 0.4261 & 0.3963 & 0.3380 & 0.2945 & +93.4\% \\
7  & \textbf{0.713} & 0.5914 & 0.5524 & 0.4343 & 0.3895 & +20.6\% & \textbf{0.830} & 0.4681 & 0.4326 & 0.3702 & 0.3173 & +77.3\% \\
8  & \textbf{0.725} & 0.6175 & 0.5861 & 0.4656 & 0.4027 & +17.4\% & \textbf{0.833} & 0.4907 & 0.4484 & 0.3854 & 0.3223 & +69.9\% \\
9  & \textbf{0.749} & 0.6473 & 0.6104 & 0.5475 & 0.4637 & +15.7\% & \textbf{0.836} & 0.5216 & 0.4869 & 0.4554 & 0.3469 & +60.3\% \\
10 & \textbf{0.774} & 0.6752 & 0.6465 & 0.5607 & 0.4965 & +14.6\% & \textbf{0.839} & 0.5216 & 0.4869 & 0.4554 & 0.3813 & +60.9\% \\
\bottomrule
\end{tabular}
\end{table*}

\begin{figure*}[t]
    \centering
    \begin{subfigure}[b]{0.49\linewidth}
        \centering
        \includegraphics[width=\linewidth]{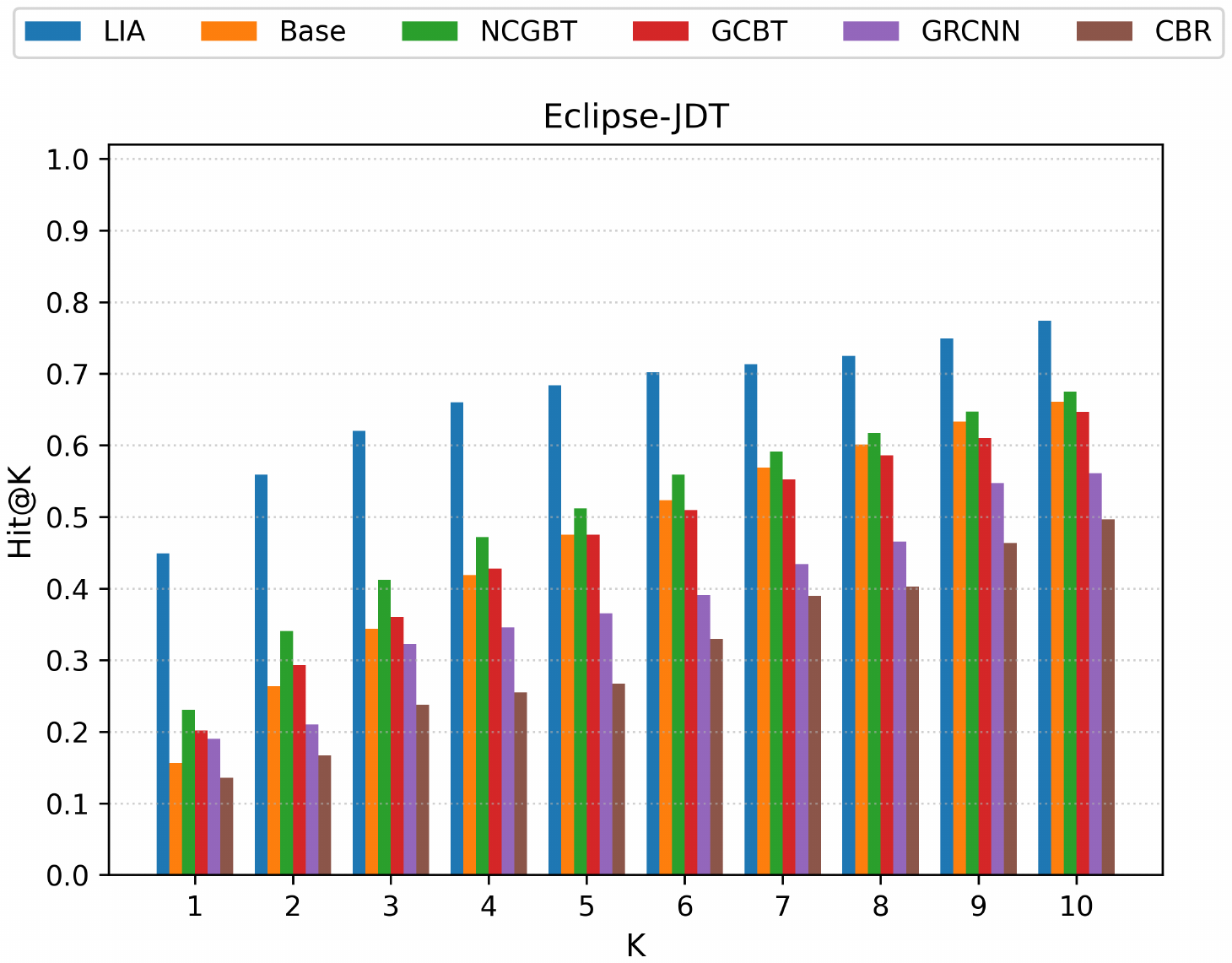}
        \caption{}
        \label{fig:a}
    \end{subfigure}
    \hfill
    \begin{subfigure}[b]{0.49\linewidth}
        \centering
        \includegraphics[width=\linewidth]{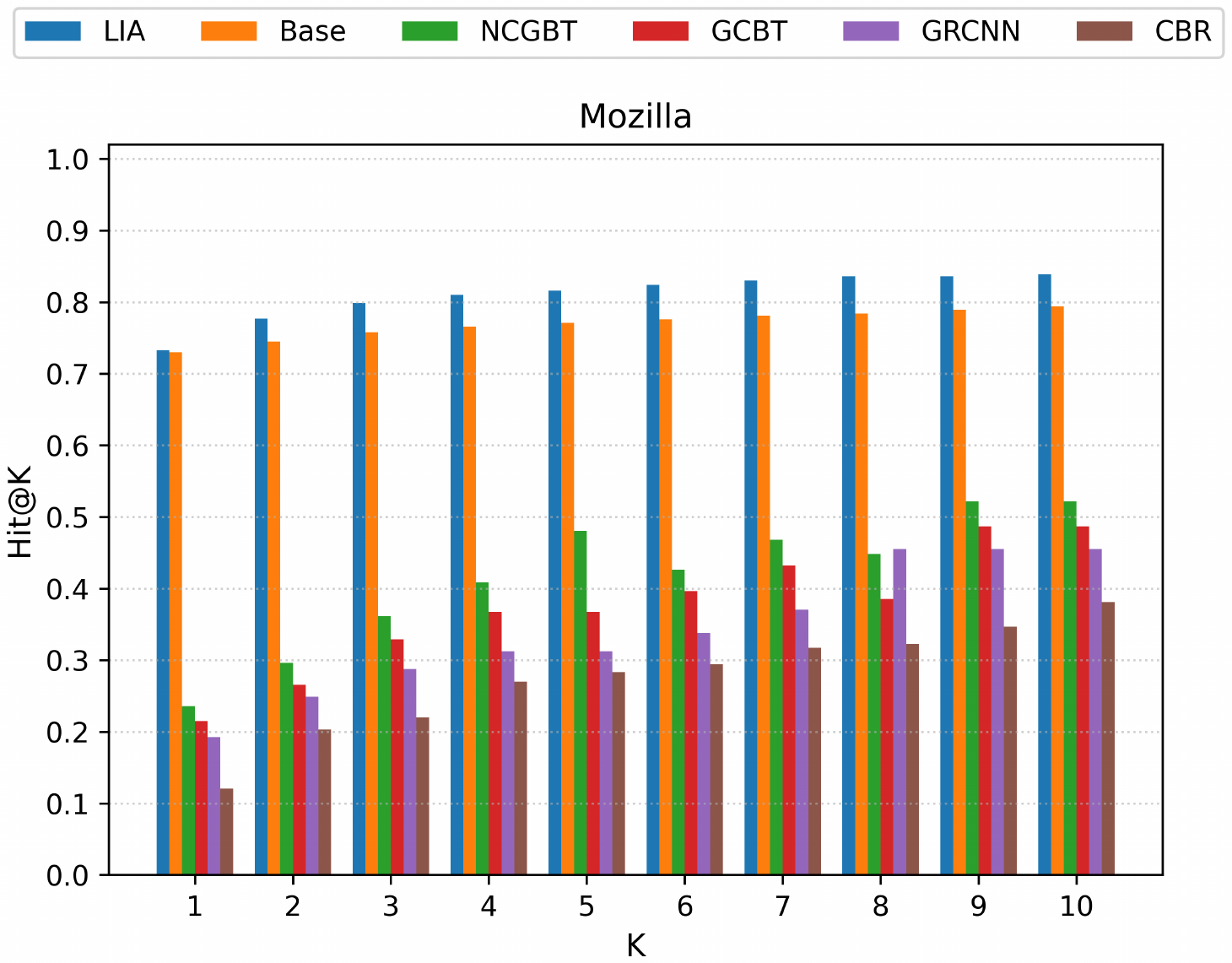}
        \caption{}
        \label{fig:b}
    \end{subfigure}
    \caption{Comparison of \SF{}, Base, and State-of-the-Art baselines (NCGBT, GCBT, GRCNN, and CBR) on the Eclipse-JDT and Mozilla}
    \label{fig:RQ2_image}
\end{figure*}

\tabref{rq2} compares \SF{} with four state-of-the-art issue assignment approaches (NCGBT, GCBT, GRCNN, and CBR) on EclipseJDT and Mozilla datasets. The ``Improve'' column in \tabref{rq2} reports \SF{}'s relative gain over NCGBT (the strongest baseline). \figref{RQ2_image} also compares \SF{}, the base model, and state-of-the-art baselines on the same two datasets. Across both datasets, \SF{} consistently outperformed all baseline methods, including the best-performing NCGBT. The improvements are particularly notable on Mozilla, where the model achieved relative gains ranging from +60.9\% at Hit@10 to +211.2\% at Hit@1. Additional gains include +162.1\% at Hit@2, +120.9\% at Hit@3, +98.3\% at Hit@4, and +99.8\% at Hit@5. On EclipseJDT, the improvements follow the same trend, starting at +14.6\% at Hit@10 and rising to +94.5\% at Hit@1, with intermediate improvements of +64.2\% at Hit@2, +50.5\% at Hit@3, +39.8\% at Hit@4, and +33.6\% at Hit@5.

One notable observation from the results is that \SF{} achieves substantially larger improvements over the baselines at lower values of $K$. In other words, the relative performance gain is most pronounced for top-ranked developer recommendations, where \SF{}'s accuracy exceeds that of competing models by the widest margins. This trend indicates that \SF{} is more effective than state-of-the-art methods at placing the correct developer near the top of the ranking. This becomes crucial in practical settings, where the highest-ranked predictions are typically the ones acted upon by project maintainers.

Another key observation is that, despite not relying on graph structures, \SF{} consistently surpassed all graph-based methods. Furthermore, \SF{} also achieved substantial gains over our text-based baseline (CBR), demonstrating that its improvements are due to its ability to capture deeper relational patterns between developers and issue contexts. In other words, \SF{} not only outperforms existing text-based approaches but also exceeds the performance of graph-based methods that currently represent the state of the art, highlighting its effectiveness in integrating semantic and contextual signals without relying on explicit graph Structures.

\begin{tcolorbox}
\textbf{Answer to RQ2:} \SF{} consistently outperformed state-of-the-art methods on both datasets, achieving improvements of Hit@1 up to +94.5\% on Eclipse-JDT and +211.2\% on Mozilla over the strongest baseline (NCGBT).
\end{tcolorbox}

%%%%%%%%%%%%%%%%%%%%%%%%%%%%%%%%%%%%%%%%%
%%
%%          Section: Discussion
%%
%%%%%%%%%%%%%%%%%%%%%%%%%%%%%%%%%%%%%%%%%
\section{Discussion}\label{sec:disc}

\paragraph{Dataset Alignment and Fine-Tuning Effectiveness}
Our results indicate that the effectiveness of supervised fine-tuning is strongly influenced by dataset-specific characteristics such as developer distribution, relational density, and data balance. On EclipseJDT, where the dataset has a smaller and denser developer set, fine-tuning produced substantial gains, particularly at lower values of $K$. In contrast, on Mozilla, the larger and more imbalanced developer pool and lower data density led to smaller yet consistent improvements. Moreover, the zero-shot performance of the pretrained model also plays a role: since the base model performed relatively poorly on EclipseJDT, there was greater room for improvement through fine-tuning, whereas on Mozilla, where the zero-shot model already achieved strong results, the potential gains were naturally more limited. Overall, these findings suggest that fine-tuning performs best when the underlying data is concentrated, and that performance variation arises from both dataset structure and the degree of alignment between the pretrained model and the target project.

\paragraph{Importance of Top-Ranked Predictions}
When comparing \SF{} with state-of-the-art methods in RQ2, the relative gains were most pronounced at smaller values of $K$. Since improvements at top ranks directly reduce manual effort and accelerate issue resolution in real-world issue assignment workflows, this trend underscores the practical importance of fine-tuning.

\paragraph{LLMs vs. Graph-Based and Text-Based Methods}
Another key finding is that \SF{} consistently outperformed state-of-the-art graph-based approaches, including NCGBT~\cite{dong2024neighborhood}. This is notable because graph-based methods have traditionally been strong baselines for issue assignment due to their ability to leverage project structure and developer–issue relationships. In addition, \SF{} also significantly outperformed our text-based baseline (CBR), showing that a fine-tuned LLM can capture deeper relational patterns between developers and issue semantics, even without relying on explicit graph structures.

%%%%%%%%%%%%%%%%%%%%%%%%%%%%%%%%%%%%%%%%%
%%
%%          Section: Threats
%%
%%%%%%%%%%%%%%%%%%%%%%%%%%%%%%%%%%%%%%%%%
\section{Threats to Validity}\label{sec:threats}
\paragraph{Internal Validity}
We relied on widely adopted open-source tools, such as the HuggingFace Hub and related Python libraries, which are well-integrated within the open-source community. However, the use of these tools also carries the risk of inheriting any existing flaws they may contain. We also acknowledge the possibility of hidden issue in our own implementation. To mitigate these risks, we performed multiple manual tests, developed a comprehensive test suite, and made both our code and data publicly available for community inspection and feedback. Our results may also be influenced by training configurations. Although we fixed random seeds and used deterministic decoding, some variability across hardware setups may still exist. Future work should systematically explore variance across random seeds, hyperparameters, and different environments to provide a more complete picture of reproducibility.  

\paragraph{External Validity}
Prior state-of-the-art evaluations have often relied on limited datasets, which constrains their generalizability. To mitigate this, we utilized two large-scale datasets representing distinct development environments; however, this scope remains limited and may still restrict the broader applicability of our findings. The positive results we observed suggest that supervised fine-tuning is effective across at least two distinct contexts, however, further studies on additional projects, domains, and industrial datasets can further assess the external validity of our approach.

\paragraph{Construct Validity}
The validity of our results depends on the performance of the underlying LLM. Since large language models exhibit a degree of stochasticity in their outputs, individual runs may show minor variations. To reduce the impact of this variability, we repeated each experiment three times and reported the average results.

\paragraph{Reproducibility} We provide the entire experiment, including the source code for \SF{} and scripts evaluation online~\cite{ReplicationPackage}. 
%%%%%%%%%%%%%%%%%%%%%%%%%%%%%%%%%%%%%%%%%
%%
%%          Section: Related Work
%%
%%%%%%%%%%%%%%%%%%%%%%%%%%%%%%%%%%%%%%%%%
\section{Related Work}\label{sec:related}

Automatic issue assignment has become an important component in modern software maintenance, aiming to reduce manual effort and speed up the resolution of reported defects~\cite{dipongkor2023comparative, lee2022light}. Over time, researchers have proposed a variety of techniques to automate this process. In this section, we group previous work into two categories: text classification-based and graph-based methods.

\subsection{Text-Based Methods}

Text-based approaches have commonly been used for issue assignment by treating parts of issue reports, such as titles and descriptions, as structured input. These inputs are turned into feature representations, and various models are trained to predict the appropriate developer as a classification label. Early work in this space used traditional ML techniques such as Naive Bayes~\cite{murphy2004automatic} and Support Vector Machines (SVM)~\cite{anvik2006automating}, coupled with methods like TF-IDF~\cite{xuan2014towards} and Latent Dirichlet Allocation (LDA)~\cite{xia2016improving}. These models, however, lacked the semantic understanding needed to capture the complex context often found in issue report text~\cite{cao2025complex, zhou2025issuecourier, zhang2023ealink, arnob2025bug}.

With the development of DL, various classification techniques have been introduced to enhance issue assignment. Wang et al.~\cite{wang2024empirical} compared 35 DL-based issue assignment models, combining different text embeddings with various classifiers. They reported that Bi-LSTM models using attention or ELMo embeddings consistently outperformed alternatives in both top-k accuracy and mean reciprocal rank (MRR). Similarly, Mani et al.~\cite{mani2019deeptriage} developed a model combining a bidirectional recurrent neural network with attention, using Word2Vec~\cite{mikolov2013efficient} to represent text and learn both syntactic and semantic features. Aung et al.~\cite{aung2022multi} introduced a multi-task learning model that simultaneously assigned developers and classified issue types. Their model used a CNN for encoding issue text and a BiLSTM for abstract syntax trees (ASTs), allowing it to incorporate structural code features often ignored by previous approaches. Arnob et al.~\cite{arnob2025bug} used XLNet for issue assignment and demonstrated that combining commit messages with summaries and descriptions led to better performance by providing richer context than text alone.

More recent studies have applied transformer models that are better at capturing contextual and semantic patterns than the mentioned DL approaches. Lee et al.~\cite{lee2022light} proposed a lightweight framework that fine-tunes pretrained transformers like RoBERTa~\cite{liu2019roberta} and DeBERTa~\cite{he2020deberta} on issue report data, achieving competitive performance with lower computational overhead. Dipongkor et al.~\cite{dipongkor2023comparative} evaluated transformer-based models across multiple open-source datasets and found that DeBERTa outperformed BERT~\cite{devlin2019bert} and RoBERTa due to its deeper attention mechanisms and disentangled representations. Their findings highlighted the ability of large pretrained transformers to capture meaningful signals without project-specific tuning.

Despite strong results, state-of-the-art text-based methods often require large amounts of labeled, project-specific data, which challenges smaller projects~\cite{zhou2025issuecourier, arnob2025bug}. They also depend on titles and descriptions whose style and availability vary across projects~\cite{akhavan2025linkanchorautonomousllmbasedagent, zhang2023ealink}, limiting generalizability. By fine-tuning a pretrained LLM we reduce the need for project-specific data, and by using standardized instruction prompts we lessen sensitivity to cross-project writing variation.

\subsection{Graph-Based Methods}

To address the limits of text-only models, recent research has explored graph-based methods that model the relationships between issues, developers, and other relevant entities. These methods are based on the idea that a developer's history of resolved issues reflects their expertise~\cite{zhou2025issuecourier, wu2022spatial, cao2025complex, dai2023graph, dong2024neighborhood}. Graph-based models build structures such as bipartite graphs or heterogeneous graphs to represent connections between reports, fixers, and code components. Graph Neural Networks (GNNs)~\cite{wu2020comprehensive} are then used to learn node embeddings that capture both the structure and content of the data.

Dai et al.~\cite{dai2023graph} proposed GCBT, which constructed a bipartite graph linking issues and developers. The model initialized issue nodes using a GRU-based text encoder and learned embeddings for both issues and developers through GNNs. It then applied an information retrieval mechanism to rank candidates. Dong et al.~\cite{dong2024neighborhood} extended this approach in NCGBT by incorporating contrastive learning, which enriched node representations by leveraging both semantic and structural neighborhood information. Tao et al.~\cite{tao2025structural} introduced SCL-BT, a self-supervised framework that uses edge perturbation and hypergraph sampling to capture both labeled and unlabeled patterns in issue–developer graphs.

Some recent work also incorporated temporal aspects of developer behavior~\cite{zhou2025issuecourier, wu2022spatial, cao2025complex}. As contributors shift roles or become inactive, capturing these changes over time becomes critical. Wu et al.~\cite{wu2022spatial} developed ST-DGNN, a model that used joint random walk sampling and recurrent graph convolution to learn patterns in developer activity at hourly, daily, and weekly scales. Zhou et al.~\cite{zhou2025issuecourier} introduced IssueCourier, a multi-relational temporal GNN that modeled five types of relationships among issues, developers, and code files, and sliced the graph over time to track evolving team dynamics. Cao et al.~\cite{cao2025complex} focused on modeling inter-issue dependencies, representing issues as nodes and blocking relationships as edges. Their method divided the graph into time-ordered snapshots and trained a ranking model incrementally to assign issues based on recent changes.

Although graph-based approaches offer strong modeling capabilities, they also present practical challenges. Many early methods struggled with scalability and high computational demands, which led researchers to adopt random walk sampling rather than full-graph learning~\cite{wu2022spatial, grover2016node2vec, nguyen2018continuous}. While this reduced complexity, it also risked missing important patterns. Additionally, these models rely on explicit links between issues and developers, which are often sparse or noisy. This reliance can degrade model performance, especially in projects with incomplete or inconsistent tracking data~\cite{tao2025structural, dong2024neighborhood}.

To address these challenges, we fine-tune a large language model (LLM) for the task of issue assignment. Leveraging the LLM's strong pretrained semantic understanding reduces the dependence on large labeled datasets. Moreover, unlike graph-based methods that require complete and consistent relational data between issues and developers, our approach operates effectively on textual information alone, thereby avoiding issues of sparsity and noise in graph construction.

%%%%%%%%%%%%%%%%%%%%%%%%%%%%%%%%%%%%%%%%%
%%
%%          Section: Conclusions
%%
%%%%%%%%%%%%%%%%%%%%%%%%%%%%%%%%%%%%%%%%%
\section{Conclusions and Future Work}\label{sec:conc}

In this paper, we introduced \SF{}, a supervised fine-tuned LLM-based approach for automated issue assignment. \SF{} leverages the pretrained semantic knowledge of an LLM and adapts it directly to the task of issue assignment. This domain adaptation enables \SF{} to capture nuanced associations between issue descriptions and developer expertise without requiring handcrafted features or project-specific graphs.
Through extensive experiments on two widely studied datasets, EclipseJDT and Mozilla, we showed that \SF{} significantly outperforms both its base pretrained model and four state-of-the-art baselines. In particular, \SF{} achieves up to +187.8\% improvements over the base model and up to +211.2\% over the strongest existing method in Hit@1 accuracy. These results establish \SF{} as a high-performing solution for large-scale issue assignment.

Our evaluation also yields important insights. First, we found that the benefits of fine-tuning are most pronounced at lower values of $K$, where accurate top-ranked recommendations matter most. Second, \SF{} was able to surpass both graph-based and text-based baselines, highlighting its ability to model developer–issue relationships without explicit graph structures. Third, the magnitude of improvement varied across datasets depending on developer distribution, data density, and the zero-shot performance of the base model, emphasizing the role of dataset alignment in fine-tuning effectiveness. Overall, these findings suggest that LLM-based models can serve as adaptable, high-precision tools for issue assignment in diverse software projects.

While the results are promising, several avenues for future research remain. First, our evaluation was limited to two datasets; extending the analysis to a broader range of projects and domains would strengthen claims about generalizability. Second, incorporating retrieval-augmented generation (RAG)~\cite{lewis2020retrieval} techniques could enable the model to dynamically access relevant historical data, such as past issue–assignee pairs or commit histories, potentially boosting performance, as RAG-based approaches have shown strong results in other software maintenance tasks~\cite{huang2025back}. Third, this study focused on fine-tuning a single LLM; future work should explore additional models and architectures to better understand their strengths and weaknesses for issue assignment. Forth, incorporating richer contextual information, such as developer profiles, project metadata, or historical collaboration patterns, into prompts could further improve assignment accuracy. Finally, beyond accuracy metrics, future research should evaluate the downstream impact of fine-tuning on developer productivity, issue resolution times, and integration into real-world workflows.
%%%%%%%%%%%%%%%%%%%%%%%%%%%%%%%%%%%%%%%%%
%%
%%          References
%%
%%%%%%%%%%%%%%%%%%%%%%%%%%%%%%%%%%%%%%%%%
%\clearpage
%%
%% The next two lines define the bibliography style to be used, and
%% the bibliography file.
\bibliographystyle{ACM-Reference-Format}
\bibliography{references}

\end{document}